\documentclass{iau} 

\usepackage{graphicx}
\usepackage{xcolor}
\usepackage[hyphens]{url}
\usepackage{hyperref}
\hypersetup{
     colorlinks  = true,
     linkcolor   = blue, 
     anchorcolor = BrickRed,
     citecolor   = blue,
     filecolor   = blue,
     menucolor   = blue,
     runcolor    = cyan,
     urlcolor    = blue
}
\usepackage[authoryear]{natbib}

\pubyear{2019}
\volume{355} 
\jname{The Realm of the Low-Surface-Brightness Universe}
\editors{D. Valls-Gabaud, I. Trujillo \& S. Okamoto, eds.}

\title[Hunting for dwarfs] 
{Hunting for low-surface
brightness features in
nearby galaxy groups}

\author[Oliver M\"uller]   
{Oliver M\"uller$^1$}

\affiliation{$^1$Observatoire Astronomique de Strasbourg (ObAS), Universite de Strasbourg – CNRS, UMR 7550, Strasbourg, France \\ email: {\tt oliver.muller@astro.unistra.fr} \\
}
\begin{document}

\maketitle

\begin{abstract}
On the scale of dwarf galaxies, several tensions between observations and the theory of structure formation have been identified in the Local Group of galaxies.  One of them, the plane-of-satellite problem describes the distribution and motion of dwarf galaxies around their hosts being planar and co-moving. To extend these studies, we have surveyed the nearby Centaurus group and found again evidence for co-rotation within a planar structure of dwarf galaxies, posing a challenge to the current $\Lambda$CDM paradigm. To further study the distribution of satellite systems around other galaxy groups, we have tested MTO -- a program to detect astronomical sources -- and found that it works well in combination with surface brightness fluctuation distance measurements to get a complete sample of dwarf galaxies. Such an approach will improve the census of dwarf galaxies in nearby galaxy groups, bringing the study of the small-scale problems to a solid statistical foundation.

\end{abstract}

\firstsection 

\section{Introduction}
In recent years, several discrepancies between the observations of dwarf galaxies and the $\Lambda$CDM model of galaxy formation have emerged. Most of these, like the missing satellite problem \citep{1999ApJ...524L..19M} or the too-big-to-fail problem \citep{2011MNRAS.415L..40B}, have been studied mainly in our own galaxy group, i.e. the Local Group. While these tensions with the $\Lambda$CDM model can be alleviated by the inclusion of baryonic physics \citep{2007ApJ...670..313S}, a more general problem arises with the distribution and motion of dwarf galaxies around their hosts. Around the Milky Way \citep{2012MNRAS.423.1109P} and the Andromeda galaxy \citep{2013Natur.493...62I}, the dwarfs seem to be aligned in planar structures which seem to co-rotate. Comparing this to $\Lambda$CDM simulations, where rather close-to isotropic distribution and random motions are expected, such observations should be very rare \citep{2014ApJ...784L...6I}. Estimations range from 0.1 to 10 percent per host galaxy to find such structures in a $\Lambda$CDM universe \citep{2012MNRAS.423.1109P,2015MNRAS.452.3838C}. This tension is now called the plane-of-satellite problem (see the recent review of \citealt{2018MPLA...3330004P}) and is extensively discussed in the literature \citep{2013MNRAS.435.2116P,2018MNRAS.477.4768B,2018A&A...614A..59B,2019ApJ...875..105P,2019arXiv190802298S,2019arXiv190907720L}. Whereas the missing satellite and too-big-to-fail problems are set on the scales of a few kiloparsec -- where indeed baryonic feedback plays a crucial role -- the plane-of-satellite problem is mainly set by the total gravity of the systems -- extending over hundreds of kiloparsec --, thus feedback within galaxies should only play a marginal role -- the baryonic solutions do not help alleviating the plane-of-satellite problem \citep{2015ApJ...815...19P}. Therefore, it is imperative to study other galaxy groups to assess whether this problem persists, or whether it is just a peculiarity of the Local Group.
The task is therefore three fold: a) search for hitherto undetected dwarf galaxies in other galaxy groups; b) measure their distances and velocities; and c) compare the groups to cosmological simulations.

\section{The Centaurus group}
We have undertaken a dwarf galaxy survey employing the Dark Energy Camera mounted at Cerro Tololo in Chile. Covering a 500 square degree field, we searched for dwarf galaxies within the Centaurus group -- consisting of Cen\,A and M\,83 -- one of the closest galaxy groups in the southern hemisphere with a mean distance of 4 Mpc. Based on their surface brightness, their sizes, their colors, and their morphology, we harvested 57 new dwarf galaxy candidates \citep{2015A&A...583A..79M,2017A&A...597A...7M}, roughly doubling the known galaxy population in this group. 

\begin{figure}
\centering
\includegraphics[width=6.5cm]{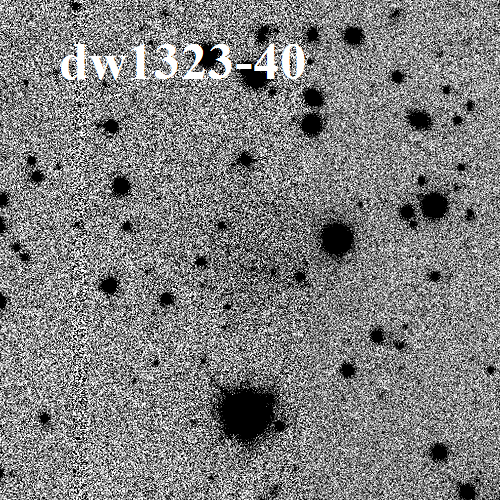}
\includegraphics[width=6.5cm]{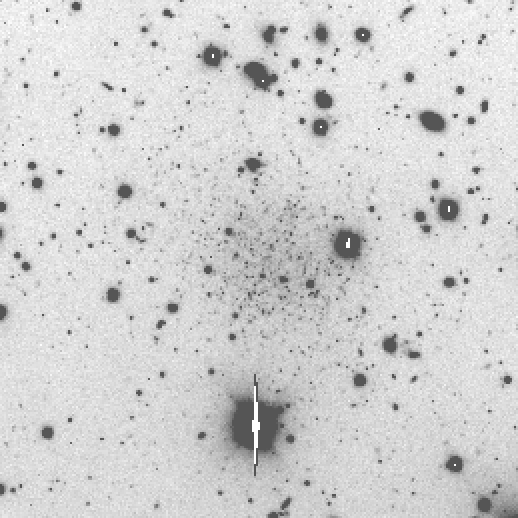}
\caption{The mugshot of dw1323-40, one of the newly discovered dwarf galaxies. Left: Discovery image with DECam, where the dwarf is barely visible \citep{2017A&A...597A...7M}. Right: The follow up with the VLT resolving the upper part of the RGB, confirming the dwarf as a member of the Centaurus group \citep{2019A&A...629A..18M}.}
\label{fig:dwarfs}
\end{figure}

To further study the Centaurus group in more detail, we have started to follow them up using FORS2 mounted at the Very Large Telescope \citep{2018A&A...615A..96M,2019A&A...629A..18M}. Under excellent seeing conditions, we have resolved the upper part of the red giant branch (RGB) of 11 dwarf galaxies in the Centaurus group and measured their distances with the tip of the RGB (TRGB). This TRGB serves as an excellent standard candle, yielding an accuracy of $\approx$5 percent. This allows us to study the three dimensional structure of Cen\,A in more detail. Combining our distance measurements with HST data by a complementary survey \citep{2016ApJ...823...19C,2019ApJ...872...80C}, we find an overall flattening of the satellite system, as it was previously suggested \citep{2015ApJ...802L..25T,2016A&A...595A.119M}. Moreover, using line-of-sight velocities for 16 dwarf galaxies, we again find signs of co-rotation around Cen\,A \citep{2018Sci...359..534M}. 
Comparing the flattening and the velocities of the dwarf galaxies to cosmological dark matter simulations yield similar results as for the Local Group planes \citep{2018Sci...359..534M}, with it appearing at the 1 percent level. This makes it the third case out of the three best studied systems to date, begging the question about their possible origins.

\section{An automatic approach to LSB detection}
Our 500 square degree survey was based on tedious visual inspection, which took months of work to scan through. While we assessed our completeness and efficiency with artificial galaxy experiments, such an approach is still prone to human error. Fortunately, there were many advances in LSB detection over the last few years. One of the most encouraging is the development of MTO \citep{teeninga2016statistical} -- a max tree based detection algorithm -- originally created for medical image analysis. We have tested this algorithm on $\approx$5 square degrees DECam data around M\,83. The exposure time was enough for measurements of surface brightness fluctuations (SBF), which is another -- less accurate than TRGB -- way to determine distances. We have scanned the image first by eye, then have applied MTO on it to produce an initial detection catalog, consisting of all significant sources in the image. By making photometric cuts in size and surface brightness, we reduced the initial detection catalog to $\approx$20 dwarf galaxy candidates. All previously known dwarfs were detected, plus a handful of new candidates. By applying SBF, we could reject all but one dwarf galaxy as background objects. The remaining detection is indeed a new ultra-faint dwarf galaxy in the Centaurus group (M\"uller et al., in preparation).
Interestingly, while in projection very close to M\,83, it is actually a member of Cen\,A and lies within the plane-of-satellites.

\section{Outlook}
While it is encouraging that the same phenomenon is found around Cen\,A, there is still a lack of data. On the one hand, multiple dwarf galaxy candidates in the Centaurus group are still missing distance and velocity information. For the latter, MUSE IFU data for an additional 14 targets around Cen\,A is now being analysed and should soon yield some results. On the other hand, there is a demand to survey other galaxy groups for dwarf galaxies. Several such surveys are now being conducted, either with public available data \citep{2017A&A...602A.119M,2018A&A...615A.105M,2017ApJ...850..109B,2019arXiv190907389C} or dedicated surveys \citep{2016A&A...588A..89J,2018ApJ...857..104G,2018ApJ...867L..15T,2018ApJ...868...96C,2019MNRAS.tmp.2031R}. Such surveys will undoubtedly uncover a plethora of dwarf galaxies, enabling us to extend our studies to new shores.

\end{document}